\documentclass[twocolumn,showpacs,preprintnumbers,amsmath,amssymb,prb]{revtex4}

\usepackage{epsfig}    
\usepackage{dcolumn}
\usepackage{bm}

\def\kp{${\bf k}\cdot{\bf p}$}
\def\eva3{$eV$$\cdot$\AA$^3$}               
\def\*#1*/{}                        

\def\ie{{\em i.e.\/}}

\newcommand{\refeq}[1]{Eq.~(\ref{#1})}
\newcommand{\refeqs}[2]{Eqs.~(\ref{#1})-(\ref{#2})}

\begin{document}

\title{Suppression of the D'yakonov-Perel' spin relaxation mechanism for all spin components in [111] zincblende quantum wells}

\author{X. Cartoix\`a}
 \email{xcs@uiuc.edu}
\affiliation{%
Department of Physics\\
University of Illinois at Urbana-Champaign\\
Urbana, IL 61801, USA
}%

\author{D. Z.-Y. Ting}%
 \affiliation{
Jet Propulsion Laboratory\\
California Institute of Technology\\
Pasadena, CA 91109, USA
}%

\author{Y.-C. Chang}
\affiliation{
Department of Physics\\
University of Illinois at Urbana-Champaign\\
Urbana, IL 61801, USA
}%

\date{\today}

\begin{abstract}
We apply the D'yakonov-Perel' (DP) formalism to [111]-grown zincblende quantum wells (QWs) to compute the spin lifetimes of electrons in the two-dimensional electron gas. We account for both bulk and structural inversion asymmetry (Rashba) effects. We see that, under certain conditions, the spin splitting vanishes to first order in $k$, which effectively suppresses the DP spin relaxation mechanism for all spin components. We predict extended spin lifetimes as a result, giving rise to the possibility of enhanced spin storage. We also study [110]-grown QWs, where the effect of structural inversion asymmetry is to augment the spin relaxation rate of the component perpendicular to the well. We derive analytical expressions for the spin lifetime tensor and its proper axes, and see that they are dependent on the relative magnitude of the BIA- and SIA-induced splittings.
\end{abstract}

\maketitle




\section{Introduction}

If the current pace of electronic device miniaturization is to continue, it is reasonable to think that the good use of the quantum properties of the electron will play a role in making this possible. Traditionally it has been the wave character of the electron that has been put to this use, resulting in devices such as the resonant tunnel diode~\cite{EsakiTsu1970} and the single electron transistor~\cite{Likharev1999}.

Another quantum property of the electron that only recently has received attention for its potential for information storage and processing is its spin. The study of spin dynamics in two-dimensional electron gases (2DEGs) is crucial to the understanding of the operation of {\em spin} elec{\em tronic} (spintronic) devices such as the Datta-Das transistor~\cite{DattaDas1990}. A newly proposed family of devices~\cite{SchliemannEguesLoss2003,CartoixaTingChang2003} is based on the special properties of the spin lifetime tensor due to the interplay between bulk inversion asymmetry~\cite{Dresselhaus1955} (BIA) and structural inversion asymmetry~\cite{BychkovRashba1984:physc} (SIA) in a [001] quantum well (QW), as pointed out by Averkiev and Golub~\cite{AverkievGolub1999}, and Kiselev and Kim~\cite{KiselevKim2000}. The effects of SIA on the spin dynamics should always be kept in mind, as it can be unintentionally present in any heterostructure due to uneven doping profiles~\cite{LuoMunekataFang1988}, surface effects, different interdiffusion at the boundaries, etc. Previous studies have found analytical expressions for the D'yakonov-Perel' (DP) spin lifetimes~\cite{DyakonovPerel1971b} in quantum wells (QWs) taking into account BIA effects only~\cite{DyakonovKachorovskii1986}, or considering both BIA and SIA for [001] QWs only~\cite{AverkievGolub1999}. There have also been numerical-only studies of the spin lifetimes in [110] QWs~\cite{LauFlatte2002}. Experimental studies in [110] structures have shown evidence of extended spin lifetimes~\cite{OhnoTerauchiAdachi1999} and lifetime modulation by the application of a gate bias~\cite{HallLauGundogdu2003}.

In this work we investigate how the interplay between SIA and BIA affects the spin lifetimes for electrons in [111] and [110] QWs. We start by computing the effective spin Hamiltonians in a two-band model for the [111] and the [110] cases. We then proceed to compute the ensemble lifetime of the three spin components as a function of the relative magnitude of BIA and SIA contributions following the procedure from Refs.~\onlinecite{PikusTitkov1984} and \onlinecite{AverkievGolubWillander2002}. We finally discuss the results and the device implications of our findings.


\section{Effective spin Hamiltonians}

To find the effective two-band Hamiltonians for the zincblende QWs, we start from the $\mathcal{O}(k^3)$ spin part of the Hamiltonian for bulk zincblendes~\cite{DyakonovPerel1971}
\begin{equation}
H_{\text{BIA}} = \gamma \left[ \sigma_x k_x \left( k_y^2 - k_z^2 \right) + c.p. \right],
\end{equation}
where $\sigma_i$ are the Pauli matrices, $k_i$ are the electron wavevector components and $c.p.$ stands for the cyclic permutation needed to obtain the remaining terms of the Hamiltonian.

We first do a change of basis to express $H_{\text{BIA}}$ in natural coordinates for the [111]- and [110]-grown structures. Then, following the procedure in Refs.~\onlinecite{DyakonovKachorovskii1986} and \onlinecite{EppengaSchuurmans1988:v37}, we quantize $\bf{k}$ along the growth direction and, keeping only terms linear in $\bf{k}_\parallel$---second order terms in $\bf{k}_\parallel$ vanish because of time reversal requirements for the expectation value of $k_z$~\cite{EppengaSchuurmans1988:v37}---, we arrive at the following expressions for the BIA Hamiltonian of [111] QWs
\begin{align}
H_{\text{BIA [111]}} &= \frac{2 \gamma \langle \hat{k}_z^2 \rangle}{ \sqrt{3} }
    \left( k_y \sigma_x - k_x \sigma_y \right),
\label{eq:BIA_Ham111} \\
\intertext{and [110] QWs~\cite{GanichevPrettl2003}}
H_{\text{BIA [110]}} &= \frac{\gamma \langle \hat{k}_z^2 \rangle}{2} k_x \sigma_z
\label{eq:BIA_Ham110}
\end{align}
where the labels $x,y,z$ depend on the orientation of the structure (see Table~\ref{tab:xyz_corresp}).

\begin{table}[t]
\centering
\begin{ruledtabular}
\begin{tabular}{cccc}
 & \multicolumn{3}{c}{Growth plane} \\
 & (001) & (110) & (111) \\ \hline
\parbox[c][4.5mm][c]{0mm}{}
 $x$ & [100] & [$\bar{1}$10] & [11$\bar{2}$] \\
\parbox[c][4.5mm][c]{0mm}{}
 $y$ & [010] & [001] & [$\bar{1}$10] \\
\parbox[c][4.5mm][c]{0mm}{}
 $z$ & [001] & [110] & [111]
\end{tabular}
\end{ruledtabular}
\caption{Correspondence between growth orientation-dependent $x$, $y$, $z$ labels and crystallographic orientations.}
\label{tab:xyz_corresp}
\end{table}

Upon inspection of the Hamiltonians in \refeqs{eq:BIA_Ham111}{eq:BIA_Ham110} we see that BIA causes a ${\bf k}$-dependent effective magnetic field pointing in-plane for [111] structures, while it points along the growth direction for [110] structures. Note that $H_{\text{BIA [111]}}$ is formally identical to the Rashba Hamiltonian~\cite{BychkovRashba1984:physc}
\begin{equation}
H_{\text{R}} = \alpha_{\text{R}}  \left( k_y \sigma_x - k_x \sigma_y \right),
\label{eq:HR}
\end{equation}
where $\alpha_{\text{R}}$ is the Rashba coefficient, whose value depends on the particulars of the structural asymmetry present in the sample. We shall now see that this has important consequences in the values of the spin lifetimes.


\section{[111]-grown structures}

We define $\alpha_{\text{BIA}} \equiv 2 \gamma \langle \hat{k}_z^2 \rangle / \sqrt{3}$. At this point we drop the [110] and [111] subindices where it is clear from the context to which structure we refer. The combination of Eqs.~(\ref{eq:BIA_Ham111}) and (\ref{eq:HR}) yields the first order Hamiltonian for [111] structures
\begin{align}
H_{\text{IA,1}} &= ( \alpha_{\text{BIA}}  +  \alpha_{\text{R}} )
  \left( k_y \sigma_x - k_x \sigma_y \right) \notag \\
              &= \alpha_{\text{IA}}  \left( k_y \sigma_x - k_x \sigma_y \right),
\label{eq:IA_111}
\end{align}
where we have introduced a parameter $\alpha_{\text{IA}} = \alpha_{\text{BIA}}  +  \alpha_{\text{R}}$ describing the combined effects of BIA and SIA in the heterostructure.

Since \refeq{eq:IA_111} is formally identical to the Rashba Hamiltonian, all the results that have been derived for the case of SIA only~\cite{AverkievGolub1999,AverkievGolubWillander2002} will also hold even when BIA is accounted for, just by making the substitution $\alpha_{\text{R}} \rightarrow \alpha_{\text{IA}}$:
\begin{equation}
\tilde{\tau}_x = \tilde{\tau}_y = 2 \tilde{\tau}_z = \frac{\hbar^2}{2 \alpha^2_{\text{IA}} }
    \frac{1}{k^2 \tilde{\tau}_1 } ,
\label{eq:111tau}
\end{equation}
where the tilde indicates a magnitude that is evaluated at a given energy and $\tilde{\tau}_1$ is the effective time for field reversal due to the harmonic $l=1$ of the scattering cross section, and in general~\cite{AverkievGolubWillander2002,LauOlesbergFlatte2001} $\tilde{\tau}^{-1}_l (E)= \oint \sigma (\phi, E) (1-\cos l \phi) d\phi$ . The spin directions will be perpendicular to the wavevector and in-plane~\cite{SchapersEngelsLange1998,CartoixaTingDaniel2001}.

Doing the thermal average of the corresponding scattering rates~\cite{AverkievGolubWillander2002} and assuming $\tilde{\tau}_1 \propto E^{\nu}$, we obtain the spin lifetimes for a nondegenerate 2DEG population,
\begin{equation}
\tau_x = \tau_y = 2 \tau_z = \frac{\hbar^2}{2 \alpha^2_{\text{IA}} }
    \frac{1}{k^2_{\theta} \tau_p } ,
\label{eq:111tau_th}
\end{equation}
\begin{equation}
\left\langle \tilde{\tau}_1 k^{2n} \right\rangle_{\theta} =
  \frac{\left( \nu + n \right) !}{\left( \nu + 1 \right) !} \tau_p k^{2n}_{\theta} \qquad n \geq 1,
\label{eq:thermal_av}
\end{equation}
with $\tau_p = \left( \nu + 1 \right) ! \; \tilde{\tau}_1 ( k_{\theta} )$ being the transport time, the symbol $\left\langle \quad \right\rangle_{\theta}$ meaning a (non-degenerate) thermal average, $k_\theta^2 = 2 m^\ast k_B T / \hbar^2$ is a thermal wavevector, $T$ is the electron temperature, $k_B$ is the Boltzmann factor, and $m^\ast$ is the electron effective mass. For the degenerate case the thermal averaging is trivial and one obtains $\tau_i = \tilde{\tau}_i \left( k_F \right)$, with $k_F$ being the Fermi wavevector. We see that, as usual in the DP mechanism, the spin lifetime is inversely proportional to the momentum lifetime.

A most interesting configuration for [111]-grown samples occurs when $\alpha_{\text{BIA}} = - \alpha_{\text{R}}$. Then, $\alpha_{\text{IA}} = 0$ and the conduction bands become spin degenerate to first order in {\bf k}. The most significant consequence of this configuration is that the spin lifetimes would be extended for {\em any} spin direction, as opposed to spins along [110] for (100) structures and $\alpha_{\text{BIA}} = \alpha_{\text{R}}$~\cite{AverkievGolub1999}, or spins perpendicular to the plane well for (110) structures and $\alpha_{\text{R}} = 0$~\cite{DyakonovKachorovskii1986}. Control of $\alpha_{\text{R}}$ can be achieved by the application of a gate bias~\cite{SchapersEngelsLange1998,NittaAkazakiTakayanagi1997} or by sample design with compositional asymmetry, providing a nonzero $\alpha_{\text{R}}$ at zero bias. Thus, properly biased (111) QWs could act as spin reservoirs, or form the basis of a resonant spin lifetime transistor as has been recently proposed for [100]~\cite{SchliemannEguesLoss2003,CartoixaTingChang2003} and [110] structures~\cite{HallLauGundogdu2003}.

\begin{figure}[t]
\centering
\epsfig{file=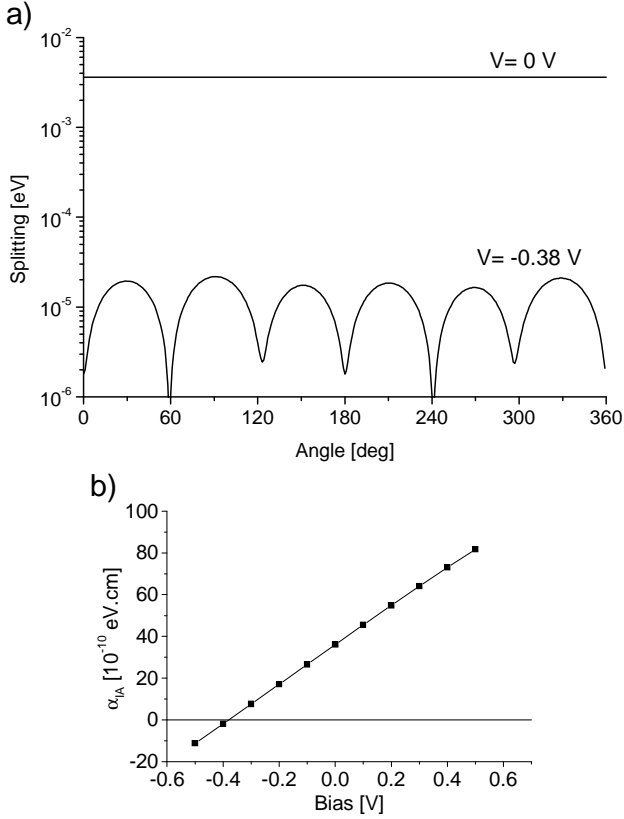, width=0.95\linewidth, clip=}
\caption{a) CB splitting as a function of angle at $k_{||} = 0.005$~\AA$^{-1}$ for an 8ML AlSb/GaSb/AlSb quantum well. Shown are results for 0~V and -0.38~V of applied bias (note the logarithmic scale). b) Value of $\alpha_{\text{IA}}$ as a function the perpendicular bias.}
\label{fig:ApplBias111}
\end{figure}

We have performed 8-band \kp\ calculations that show the validity of our treatment. Figure~\ref{fig:ApplBias111} shows the conduction band spin splitting as a function of the angle of ${\bf k}_{||}$ with respect to $k_x$ for zero applied perpendicular bias and -0.38~V, where we find that $\alpha_{\text{BIA}} \approx - \alpha_{\text{R}}$. It is clearly seen that for $V=0$~V (\ie\ $\alpha_{\text{R}}=0$) the splitting is fairly independent of the angle, an indication of the predominance of the linear regime. On the other hand, for $V=-0.38$~V we see a much smaller splitting (note the logarithmic scale) with an oscillatory character indicative of higher order in {\bf k} terms dominating the Hamiltonian [see \refeq{eq:111O3} below]. The inset shows the calculated dependence of $\alpha_{\text{IA}}$ (through $\alpha_{\text{R}}$) on the voltage applied, illustrating the control of $\alpha_{\text{IA}}$ with a gate voltage. We believe that the high values of $\alpha_{\text{BIA}}$ are due to [111] being the direction of maximum asymmetry in the zincblende structure.

We can find the limitation that higher order terms impose upon these resonant spin lifetimes by following the procedure of Averkiev {\em et al.}~\cite{AverkievGolubWillander2002}. For (111) structures, the effective CB Hamiltonian will be
\begin{multline}
H_{\text{IA,3}} = \frac{\gamma}{2\sqrt{3}} \left[ k^2
    \left( -k_y \sigma_x +
    k_x \sigma_y \right) + \right. \\
    \left. \sqrt{2} \left( 3 k_x^2 - k_y^2 \right) k_y \sigma_z
    \right] .
\label{eq:111O3}
\end{multline}
If we add \refeq{eq:111O3} to \refeq{eq:IA_111}, we obtain the following results for the spin lifetimes
\begin{align}
\tilde{\tau}_x &= \tilde{\tau}_y = \frac{6 \hbar^2}{k^2 \tilde{\tau}_1}
    \frac{1}{12 \alpha_{\text{IA}}^2 - 4 \sqrt{3} \gamma \alpha_{\text{IA}} k^2 +
        (1 + 2 \tilde{\tau}_3 / \tilde{\tau}_1 ) \gamma^2 k^4 } \notag \\
\tilde{\tau}_z &= \frac{3 \hbar^2}{k^2 \tilde{\tau}_1}
    \frac{1}{ \left( \gamma k^2 - 2 \sqrt{3} \alpha_{\text{IA}} \right)^2 } .
\label{eq:tau111O3}
\end{align}

\begin{figure}[t]
\centering
\epsfig{file=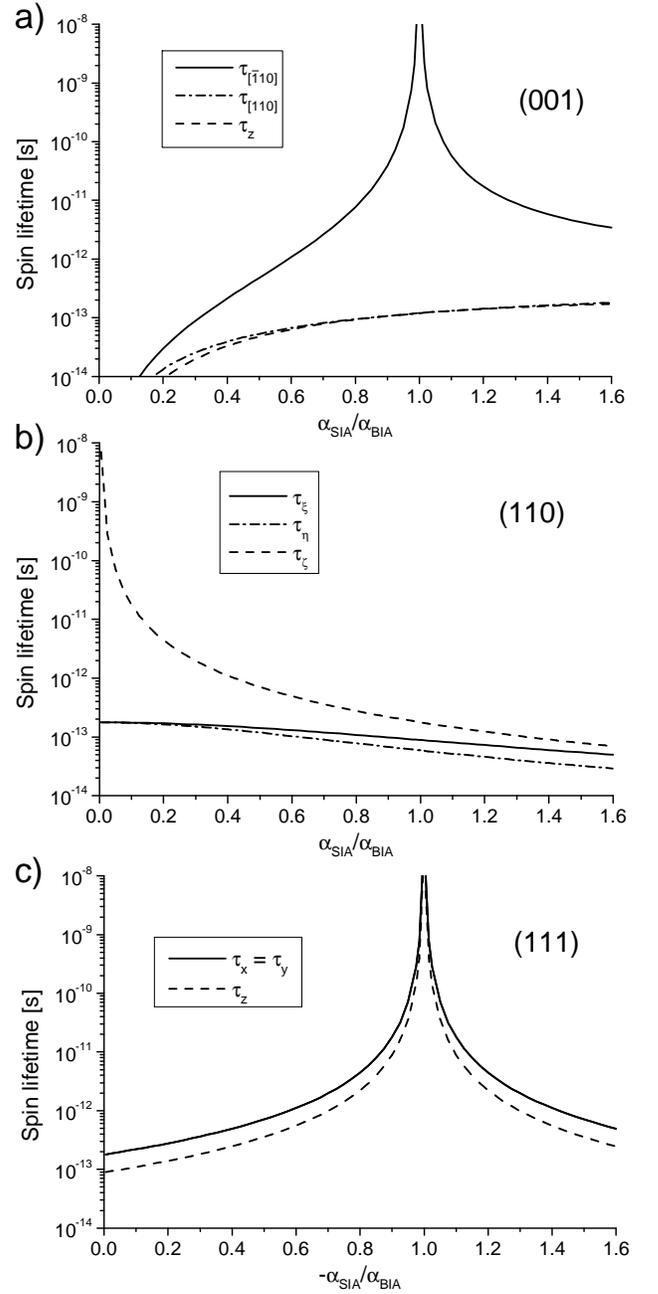, width=0.95\linewidth, clip=}
\caption{Spin lifetimes for the three spin components for \mbox{[001]-,} [110]- and [111]-grown QWs as a function of the ratio of the SIA and the BIA parameters. Note the resonance for all components for in the case of a [111] structure, as opposed to a single component resonance for [001] and [110] structures. The labels $\tau_{\xi,\eta,\zeta}$ for the [110] orientation refer to the spin relaxation tensor proper axes as defined in \refeq{eq:tau3}.}
\label{fig:lifetimes}
\end{figure}

Since the scattering rate is proportional to $(H_{\text{IA}})^2$, the $k^6$ terms in \refeq{eq:tau111O3} are not correct in general because terms arising from the combination of $H_{\text{IA,1}}$ with fifth order contributions to $H_{\text{IA}}$ are missing. However, we have kept the $k^6$ terms here because they are correct in the special case where $\alpha_{\text{IA}} = 0$, giving the lowest order contribution to the spin scattering rate.

Figure~\ref{fig:lifetimes} shows the calculated lifetimes in the present treatment for the two structures under study, and previous [001] results~\cite{AverkievGolub1999,CartoixaTingChang2003} are provided for convenience. Equation~(\ref{eq:tau111O3}) is plotted in Fig.~\ref{fig:lifetimes}.a) for typical values $k_F=0.01$ \AA$^{-1}$, $\tau_p=1$ ps, $\gamma=186$ eV$\cdot$\AA$^{3}$~\cite{CardonaChristensenFasol1988}, $\alpha_{\text{BIA}}=11 \times 10^{-10}$ eV$\cdot$cm~\cite{CartoixaTingMcgill2003b} as a function of the ratio $\alpha_{\text{R}} / \alpha_{\text{BIA}}$. The three spin lifetime components show the predicted resonant spin lifetime when $\alpha_{\text{R}} = -\alpha_{\text{BIA}}$. \refeq{eq:tau111O3} by itself limits the lifetimes at the resonances to finite but very large values. In practice, other mechanisms~\cite{Elliott1954,Yafet1963,BirAronovPikus1975} will effectively limit the value of the resonant spin lifetime. Therefore, [111]-grown heterostructures provide DP suppression on par with [110]-grown structures, with the added advantage the suppression is for {\em all} spin components, as opposed to one component only.


\section{[110]-grown structures}

We also apply the present formalism to derive analytical expressions for the spin lifetimes in [110] QWs under the combined action of BIA and SIA. Here we define $\alpha_{\text{BIA}} \equiv \gamma \langle \hat{k}_z^2 \rangle / 2$, and the effective Hamiltonian due to inversion asymmetry is
\begin{equation}
H_{\text{IA,1}} = \alpha_{\text{BIA}} k_x \sigma_z  +  \alpha_{\text{R}}  \left( k_y \sigma_x - k_x \sigma_y \right),
\label{eq:IA_110}
\end{equation}
which induces a spin splitting $\Delta_{\text{IA,1}}$ equal to
\begin{equation}
\Delta_{\text{IA,1}} = 2 k \sqrt{\alpha_{\text{R}}^2 + \alpha_{\text{BIA}}^2 \cos^2 \phi},
\label{eq:sp_110}
\end{equation}
where $k$ and $\phi$ are defined by $k_x=k \cos \phi$ and $k_y=k \sin \phi$.

\begin{figure}[t]
\centering
\epsfig{file=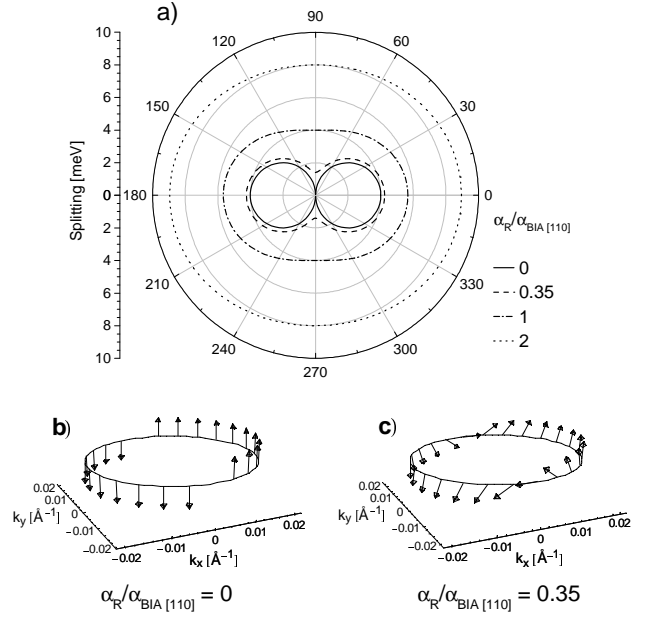, width=0.95\linewidth, clip=}
\caption[Spin splitting and directions for the lower CB of a 110-grown QW.]{Spin splitting and directions for the lower conduction subband of a [110]-grown QW. Plot a) shows the value of the spin splitting from \refeq{eq:sp_110} for states lying on a circle in the $k_x-k_y$ plane with $k=0.01$ \AA$^{-1}$ for various values of the ratio $\alpha_{\text{R}} / \alpha_{\text{BIA}}$. Plots b) and c) show the spin eigenstates for \refeq{eq:IA_110} for $\alpha_{\text{R}} / \alpha_{\text{BIA}} = 0$ and $\alpha_{\text{R}} / \alpha_{\text{BIA}} = 0.35$, respectively. Note that when $\alpha_{\text{R}} = 0$ all spins are aligned with the $z$ axis. In that case, no spin direction is specified for the $[010]$ and $[0\bar{1}0]$ directions because the states are spin degenerate. The value assumed for $\alpha_{\text{BIA}}$ is $10^{-9}$ eV$\cdot$cm.}
\label{fig:spin_split_110}
\end{figure}

Figure~\ref{fig:spin_split_110} shows the spin splitting and directions for the lower conduction band of a [110]-grown QW for different values of $r \equiv \alpha_{\text{R}} / \alpha_{\text{BIA}}$ and $k$ fixed to 0.01~\AA$^{-1}$. In plot a), the distance from the line to the center indicates the amount of splitting, and the angle corresponds to $\phi$. For $r=0$ we see that spin splitting is highly anisotropic, coming from the BIA contribution of \refeq{eq:IA_110}. As the SIA effects, which are inherently isotropic, grow, the splitting becomes more and more isotropic. The spin directions for $r=0$ [plot b)] have very interesting behavior. The effective magnetic field always points along $\pm z$, independently of ${\bf k}$~\cite{DyakonovKachorovskii1986}. This will result in absence of precession for spins along $z$, and the DP mechanism for spin relaxation will be suppressed for that component of spin~\cite{DyakonovKachorovskii1986}.

The proper axes of the spin lifetime tensor are dependent on the value of $\alpha_{\text{R}}$ and $\alpha_{\text{BIA}}$. Their orientation $i$ and their corresponding lifetimes $\tilde{\tau}_i$ are found by diagonalizing the spin scattering rate tensor:
\begin{align}
\tilde{\tau}_\xi &= \frac{\hbar^2}{2 \left( \alpha^2_{\text{R}} + \alpha^2_{\text{BIA}} \right) }
    \frac{1}{k^2 \tilde{\tau}_p } &  
    \xi \rightarrow \hat{x}
 \notag \\
\tilde{\tau}_\eta &= \frac{\hbar^2}{2 \left( 2 \alpha^2_{\text{R}} + \alpha^2_{\text{BIA}} \right) }
    \frac{1}{k^2 \tilde{\tau}_p } &  
    \eta \rightarrow
    \frac{\alpha_{\text{BIA}} \hat{y} + \alpha_{\text{R}} \hat{z} }{ \sqrt{\alpha^2_{\text{R}} + \alpha^2_{\text{BIA}} } } \notag \\
\tilde{\tau}_\zeta &= \frac{\hbar^2}{2 \alpha^2_{\text{R}} }
    \frac{1}{k^2 \tilde{\tau}_p } &  
    \zeta \rightarrow
    \frac{-\alpha_{\text{R}} \hat{y} + \alpha_{\text{BIA}} \hat{z} }{ \sqrt{\alpha^2_{\text{R}} + \alpha^2_{\text{BIA}} } } 
\label{eq:tau3}
\end{align}

Equation~(\ref{eq:tau3}) is plotted in Fig.~\ref{fig:lifetimes}.b) for the same values as in Fig.~\ref{fig:lifetimes}.a). The $\tau_\xi$ and $\tau_\eta$ components show regular behavior, while the $\tau_\zeta$ component shows the resonant spin lifetime when $\alpha_{\text{R}} = 0$~\cite{DyakonovKachorovskii1986,SchapersEngelsLange1998} that has been justified above. If $\alpha_R$ is due to some external bias, we see from \refeq{eq:tau3} that the spin lifetime along direction $\zeta$ will decrease proportionally to $E^{-2}$, where $E$ is the applied electric field perpendicular to the QW plane. In practice, this resonance will have a finite peak height due to the other mechanisms~\cite{Elliott1954,Yafet1963,BirAronovPikus1975} starting to kick in. In principle, higher order in {\bf k} terms might also limit the spin lifetime at the resonance. However, we see that the $k^3$ terms that should be added to \refeq{eq:IA_110} are given by
\begin{equation}
H_{\text{IA}}^3 = \gamma \left( -k_x^2 /2  + k_y^2 \right) k_x \sigma_z .
\label{eq:110O3}
\end{equation}
The direction of the effective magnetic field is still $z$ irrespective of the value of {\bf k}. As a consequence, the resonance is not destroyed by the inclusion of the $k^3$ terms, as noted by Hall {\em et al.}~\cite{HallLauGundogdu2003}. As for the case of [111] structures, the possible effect of higher than $k^3$ terms in the Hamiltonian would be shadowed by the onset of the other spin relaxation mechanisms.

\section{Summary}

In conclusion, we have calculated and given analytical expressions for the electron spin lifetime for [111] and [110] QWs including both BIA and SIA (Rashba) effects. We find that the D'yakonov-Perel' (DP) spin relaxation mechanism can be effectively suppressed for {\em all} spin components in [111] QWs at a resonance condition through appropriate sample design or the application of a suitable gate bias. The $z$ spin component of [110] QWs can be made to show the same level of suppression of the DP mechanism, but the other two components will always have the non-resonant dynamics. This effect in [111] QWs can lead to the relaxation of design constraints in devices such as the spin-LED or the resonant spin lifetime transistor. Also, these structures may act as a spin storage stage.

\acknowledgements
This work was supported in part by Defense Advanced Research Projects Agency (DARPA) under Contracts No. MDA972-01-C-0002 and No. DAAD19-01-1-0324. A part of this work was carried out at the Jet Propulsion Laboratory, California Institute of Technology, through an agreement with the National Aeronautics and Space Administration.

\bibliographystyle {apsrev}
\bibliography{../../bibtex/complete}

\end{document}